# Estimating the relative contribution of streetlights, vehicles and residential lighting to the urban night sky brightness

**Salvador Bará,**[1,*] **Ángel Rodríguez-Arós,**[2] **Marcos Pérez,**[3] **Borja Tosar,**[3] **Raul C. Lima,**[4,5] **Alejandro Sánchez de Miguel,**[6] **and Jaime Zamorano**[7]

[1] *Facultade de Óptica e Optometría, Universidade de Santiago de Compostela, 15782 Santiago de Compostela, Galicia, Spain*

[2] *ETS Náutica e Máquinas, Universidade da Coruña, Galicia, Spain*

[3] *Casa das Ciencias, Museos Científicos Coruñeses, A Coruña, Galicia, Spain*

[4] *Escola Superior de Saúde, Politécnico do Porto (ESS/PPorto), Portugal*

[5] *CITEUC, Centre for Research on Earth and Space of the University of Coimbra, Portugal*

[6] *Instituto de Astrofísica de Andalucía IAA-CSIC, Granada, Spain*

[7] *Dept. de Astrofísica y CC. de la Atmósfera, Fac. de Ciencias Físicas, Universidad Complutense, Madrid, Spain.*

[*] *Corresponding author: salva.bara@usc.es*

**Abstract**

Under stable atmospheric conditions, the zenithal brightness of the urban sky varies throughout the night following the time course of the anthropogenic emissions of light. Different types of artificial light sources (e.g. streetlights, residential, and vehicle lights) present specific time signatures, and this feature makes it possible to estimate the amount of sky brightness contributed by each one of them. Our approach is based on transforming the time representation of the zenithal sky brightness into a modal coefficients one, in terms of the time course signatures of the sources. The modal coefficients, and hence the absolute and relative contributions of each type of source, can be estimated from the measured brightness by means of linear least squares fits. A method for determining the time signatures is described, based on wide-field time-lapse photometry of the urban nightscape. Our preliminary results suggest that artificial light leaking out of the windows of residential buildings may account for a significant share of the time-varying part of the zenithal sky brightness, whilst the contribution of the vehicle lights seems to be significantly smaller.

*Keywords: Artificial light at night, light pollution, radiometry, photometry*

## 1. Introduction

Light pollution is a widespread phenomenon in the modern world [1-4]. In many places of our planet the natural levels of night darkness are being significantly disrupted, an unwanted side-effect of the progressive extension of the use of artificial light [5-8]. One of the most conspicuous manifestations of this problem is the increased sky glow that hinders the observation of the starry sky and acts as an additional light source that increases the radiance and irradiance levels at otherwise pristine natural sites, located many kilometers away from the urban conglomerates [1-4, 9-14].

Several types of artificial sources contribute to this phenomenon. First and foremost the outdoor public and private lighting systems improperly designed or installed, that leak a relevant fraction of light towards the sky either directly or after reflecting off the ground unnecessarily high illuminance levels. Vehicle lights also contribute to the overall sky glow: some fraction of the energy emitted by the headlights propagates at angles above the horizontal, and a non-negligible amount is reflected by the pavement, with a combined Lambertian-directional angular pattern that depends on the pavement roughness and on the prevailing atmospheric conditions (in particular, rain). Finally, commercial and residential indoor lights, leaking out of the windows, also contribute with complex radiance patterns to the overall sky brightness.

Estimating the relative weights of these contributions is an issue of interest not only for basic science, but also for practical policy-making. A significant reduction of the present light pollution levels can only be achieved by acting on the sources, and the approaches and timing of any practicable corrective measures will necessarily be different for each of them. Whilst most efforts have been devoted to re-design in a more efficient way the static outdoor lighting systems, a reappraisal of the present design of vehicle headlights and some cultural changes in the way window shutters and blinds are used in some cities could help to achieve additional reduction goals.

The estimation of the relative contribution of each type of source to the overall sky brightness can be made using two different approaches. One of them is based on carrying out a detailed census of the individual light emitting points (streetlights, lighted windows of homes and shops, vehicles,...) and evaluating their contributions to the night sky brightness by using suitable theoretical models and software tools that allow to calculate the amount of light scattered back from the atmosphere in the direction of the observer [15-23], taking into account the individual radiance patterns of each source (spectral and angular), the ground and wall reflections undergone by the light before being redirected towards the sky, and the prevailing meteorological conditions.

A second, complementary approach, that we present in this paper, is to determine directly these contributions by making a global estimation based on the characteristic time course signature of each type of artificial source. Considered as a whole, and averaged over the city area, the streetlights, vehicles, and domestic lights contribute to the overall sky glow with characteristic time-varying patterns. As we show in this paper, the overall sky brightness, measured in radiant (Wm$^{-2}$sr$^{-1}$) or in light (cd/m$^2$) units, can be written as a linear combination of these specific time course functions, and a linear least-squares estimation of the relative weight of each term can be easily made if the overall sky brightness is measured with sufficient time resolution and the shape of the time signature functions is known, either by direct measurement or through indirect methods. The results from this analysis provide the absolute and relative contributions of each type of source to the overall sky brightness at any time of the night, and can be a complementary aid for making informed decisions on public lighting policies and modifying, if needed, the prevailing forms of using artificial light and night.

The structure of this paper is as follows. In Section 2 we briefly introduce the basis of the global time course signature approach. Section 3 describes the materials and methods we used to acquire and process the observational data gathered in the campaigns described in Section 4. A discussion of the limitations and significance of this work is carried out in Section 5, and conclusions are drawn in Section 6. Some additional formalization can be found in the Appendix.

## 2. Estimating the components of the time course of the urban night sky brightness

In clear and moonless nights, under constant atmospheric conditions, the zenithal sky brightness in strongly light-polluted cities varies throughout the night following the time course of the artificial emissions of light. Assuming that $M$ different kinds of sources (e.g. streetlights, industrial, vehicle, ornamental, domestic lights, …) contribute to the overall sky glow, the time course of the zenithal night sky brightness $B(t)$, expressed in weighted radiance (Wm$^{-2}$sr$^{-1}$) or luminance (cd/m$^2$) units, can be written as:

$$B(t) = \sum_{i=1}^{M} c_i T_i(t) \qquad (1)$$

where $T_i(t)$, $i = 1,...,M$, are a set of functions that describe the time course of the emissions of the different types of light sources, and $c_i$ are constant factors that account for the relative weight of each type of source in the total zenithal sky brightness (see Appendix for formal details). The normalization and units of $c_i$ and $T_i(t)$ can be freely chosen, as long as their product in Eq. (1) has the proper value in radiance or luminance units. A practical choice, among other possible options, is to normalize $T_i(t)$ such that it is dimensionless and equal to 1 at initial time, that is, $T_i(t=0)=1$. According to Eq. (1), the relative contribution of the *k*-th type of source to the overall zenithal sky brightness at time $t$, $\gamma_k(t)$, is given by:

$$\gamma_k(t) = \frac{c_k T_k(t)}{\sum_{i=1}^{M} c_i T_i(t)} \qquad (2)$$

From Eq. (2) the relative contributions $\gamma_k(t)$ at any time of the night can be determined, if the values of the constants $c_i$ and the time signature functions $T_i(t)$, $i = 1,...,M$, are known. The $c_i$ constants, in turn, can be estimated in a least-squares sense from the measured values of the zenithal sky brightness, $B(t)$, and of $T_i(t)$. Let us assume that these values are available for a sufficiently large array of time points $t_s$, $s = 1,...,N$, where $N$ is the total number of measurements of each function. Since Eq. (1) applies individually for each $t_s$, it can be rewritten in matrix-vector form as:

$$\mathbf{b} = \mathbf{T}\mathbf{c} \qquad (3)$$

where $\mathbf{b}$ is a vector of size $N \times 1$, with components $b_s = B(t_s)$, $\mathbf{T}$ is a matrix of size $N \times M$ whose elements are given by $T_{sk} = T_k(t_s)$, and $\mathbf{c}$ is a $M \times 1$ vector whose components are the unknown coefficients $\{c_k\}$. If the number of measurements exceeds the number of unknowns, i.e. if $N > M$, the linear system in Eq. (3) is overdetermined and can be solved, in the least-squares sense, by [24]:

$$\hat{\mathbf{c}} = \mathbf{T}^+ \mathbf{b} \qquad (4)$$

where $\mathbf{T}^+$ is a suitable pseudoinverse matrix, as e.g. $\mathbf{T}^+ = (\mathbf{T'T})^{-1}\mathbf{T'}$, with $\mathbf{T'}$ standing for the transpose of $\mathbf{T}$. The symbol ^ indicates that $\hat{\mathbf{c}}$ is an estimate of the true vector $\mathbf{c}$: these vectors are not expected to be strictly coincident, due to the unavoidable propagation to $\hat{\mathbf{c}}$ of the noise present in the measurements of $B(t)$ and, also, in the elements of $\mathbf{T}$, if they are determined experimentally as it is our case and not from theoretical first principles. Finally, by substituting the values of the elements of $\hat{\mathbf{c}}$ for those of $\mathbf{c}$ into Eq. (2) we get the estimated contribution of each type of light source to the zenithal sky brightness.

## 3. Materials and Methods

### *3.1. Measuring the zenithal night sky brightness*

We measured the zenithal night sky brightness, $B(t)$, using two different models of low-cost SQM light meters (Unihedron, Canada). These devices are based on a TSL237 (TAOS, USA) high-sensitivity irradiance-to-frequency converter, with temperature correction, fitted with optics that restricts its field of view to a region of the sky with an approximately Gaussian weighting profile and full-width at half maximum 20° [25]. The spectral band is limited to 400-650 nm (effective passband of the whole setup, at half the maximum sensitivity) [25-26].

The readings of the SQM light meters are given in the negative logarithmic scale of *magnitudes per square arcsecond* (mag/arcsec$^2$), a non-SI unit for the integrated spectral radiance, scaled and weighted by the spectral sensitivity of the detector [14]. An offset of 0.1 mag/arcsec$^2$ shall be subtracted from the raw measurements, to account for the losses at the glass window of the detector housing, mostly due to Fresnel reflections at the air-glass interfaces. Due to the specific spectral sensitivity of the SQM device, the conversion between the SQM magnitudes and the corresponding mag/arcsec$^2$ in other standard photometric bands, like e.g. the ones defined by the Johnson-Cousins B, R or V filters [27], or the CIE photopic visual spectral efficiency function V($\lambda$) [28], can only be done in an approximate way if *a priori* spectral information is not available [29]. The V-band radiance $L_V$ (in Wm$^{-2}$sr$^{-1}$) and the corresponding value of the $m_V$ magnitude in units mag$_V$/arcsec$^2$ are related by [14]:

$$L_V [W\cdot m^{-2} sr^{-1}] = 158.1 \times 10^{(-0.4 m_V)} . \qquad (5)$$

Under the (only approximate) assumption that visible V($\lambda$) luminances can be directly estimated from Jonson-Cousin V radiances, the luminance $L$ in cd/m$^2$ (equivalent to lx/sr) can be calculated by multiplying Eq.(5) by the 683 lm/W luminous efficacy factor, obtaining the standard conversion formula [4,14-15]:

$$L[cd\cdot m^{-2}] = 10.8 \times 10^4 \times 10^{(-0.4 m_V)} . \qquad (6)$$

In our measurements, the zenithal night sky brightness $B(t)$ was continuously monitored throughout the night, at rates ranging from one reading every 42 s to one reading per minute, depending on location. These rates are considerably faster than the characteristic times in which the contributions of most types of artificial light sources, aggregated at the city level, are expected to vary; this oversampling, however, provides a high number of data that are instrumental for attenuating the effects of the measurement noise.

### 3.2. Determining the time course of the artificial sources of light

The relevant types of artificial light sources may vary from site to site, and shall be determined after analyzing the specific features of each urban area. Streetlights, vehicle headlights and light leaking out from the windows of residential buildings are three relevant sources that shall be taken into account in most cities and towns. Other sources that may contribute significant amounts of sky glow with specific time courses, as e.g. industrial facilities or ornamental lighting, shall also be included when relevant.

In this work we estimated the time course of the artificial components, $T_k(t)$, by means of wide-field time-lapse photometry of the urban nightscape, using off-the-shelf digital single-lens reflex cameras (DSLR) [30-34]. This technique has been successfully used by Dobler et al (2015) to determine the aggregated behaviour of residential and commercial light sources in New York [35]. Here we extend this approach to the traffic flow and to some specific source types, as outdoor industrial or ornamental lighting. The basic procedure consists in acquiring from a fixed vantage point successive images of the urban nightscape, containing the basic types of sources under study (Fig. 1), and extracting from these images the temporal evolution of the source radiances.

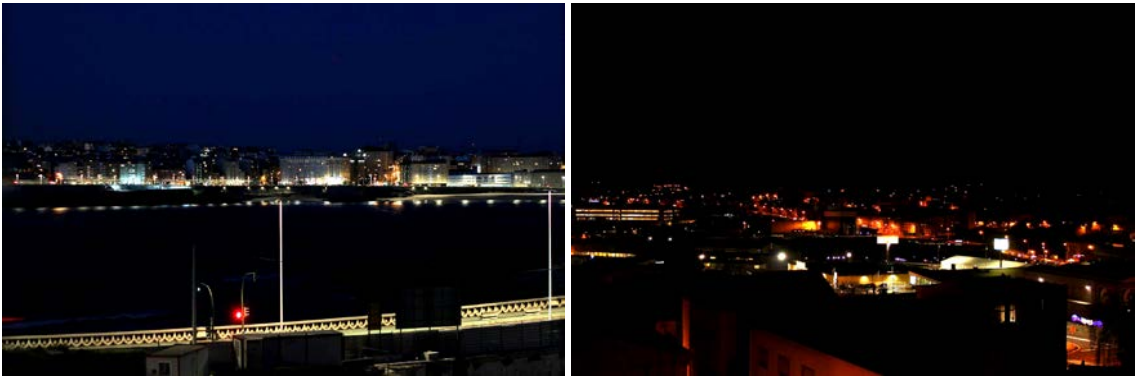

**Figure 1**: Examples of individual time-lapse frames taken at the two sites analyzed in our study, A Coruña (left) and Arteixo (right), both located in Galicia (Spain)

The images are taken at typical rates of one frame per minute, and stored in RAW format for subsequent off-line analysis. The focal length of the camera lens is chosen to attain the desired field of view, and the ISO sensitivity, diaphragm aperture and exposure times are settled to maximize the use of the available dynamic range of the CCD or CMOS camera sensor. Details of the parameters used in our measurements can be found in Section 4.

Our time-lapse analysis workflow proceeds as follows. First, the images in the RAW proprietary camera format are converted to the open, loss-less, DNG RAW format using a free application as e.g. Adobe's DNG converter [36]. The resulting DNG files are read with MATLAB^TM (MathWorks, USA), and processed to extract the Bayer color filter array (CFA), linearize it, apply the appropriate white balance multipliers, and interpolate each color channel in order to get a full-resolution RGB image (demosaicing). This RGB image is then converted into graylevel luminance by means of a suitable linear combination of its RGB color channels.

A set of binary masks is constructed, each one corresponding to the regions of the image that contain a given type of source (e.g. residential buildings windows, or vehicle traffic roads). The time-lapse luminance images are successively multiplied by each mask and the average value of the pixels corresponding to each region is calculated. The outcome is a set of linear data arrays containing the average pixel values at each time of the night and for each region. Then, the constant background light is subtracted from the arrays whose binary mask contained pixels not belonging to the sources themselves (e.g. segments of façades with no windows, or road surfaces in periods of no traffic), and all arrays are normalized such that their initial element is equal to 1. The resulting arrays are taken as an estimate of the values of $T_k(t_s)$ for $k=1,...,M$ and $s=1,...,N$. As a final step, and to ensure working in an homogeneous time frame when applying Eq. (4), the SQM luminances (see section above) are linearly interpolated between measurements at times coincident with the time-lapse $t_s$ array.

4. Results

A proof-of-concept of this method was carried out in two cloudless and moonless nights in the cities of A Coruña (250.000 inh.) and Arteixo (30.000 inh.), located in the Atlantic shoreline in the North West coast of Galicia (Spain). The parameters and results of each observation night are given below. In both cases the analysis was restricted to the period comprised between the end of the evening astronomical twilight and the beginning of the morning astronomical twilight, that is, the astronomical night (Sun altitude below −18°).

*4.1. A Coruña*

The measurements in A Coruña were made in the night of 27 to 28 April, 2017. The zenithal sky brightness was measured from the premises of the *Casa das Ciencias Planetarium* (*Museos Científicos Coruñeses*) located in the center of the city, using a SQM-LU detector taking one reading every 42 s. The time course of the components of the artificial light emissions was determined by time-lapse photometry using the DSLR images or the urban nightscape recorded from the *ETS de Náutica e Máquinas* of the University of A Coruña, located 1.1 km to the NNW of the former location (Fig 1, left). These images were acquired at a rate of one frame per minute with a 50 mm fixed focal length lens set at f/2.8 attached to a Canon EOS 1200D Digital SLR standing on a tripod and set to ISO 400 and shutter speed of 1/10 s. Since the emissions of the outdoor public lighting system of A Coruña municipality are known to be kept at a constant level throughout the whole night, only the residential and vehicle lights components were determined from these images. They were subsequently normalized as described in Section 3.2. The resulting $T_k(t)$ functions, as well as the SQM radiances normalized to 1 at the initial time point, are shown in Fig. 2.

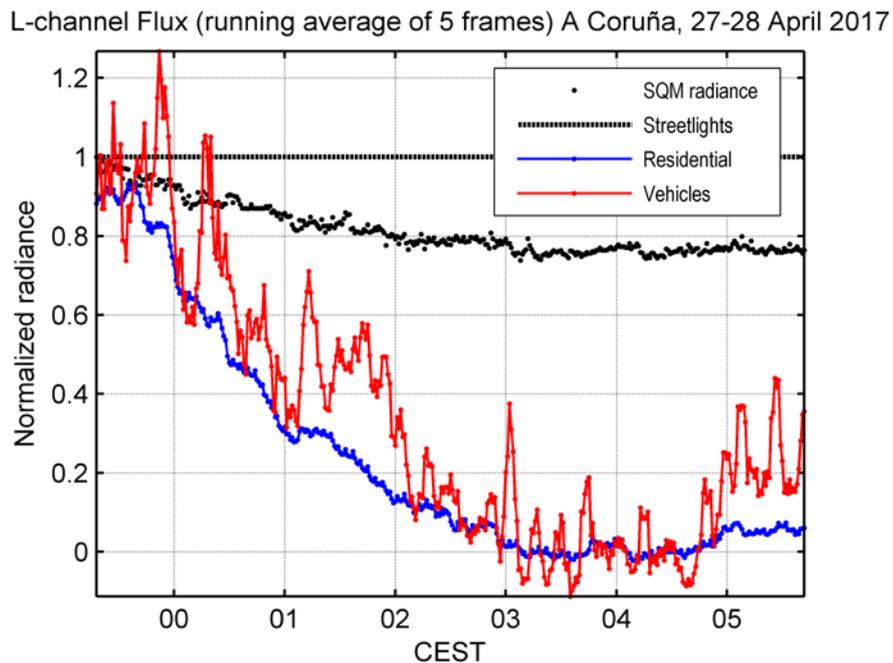

**Figure 2**: Normalized time course $T_k(t)$ of the zenithal night sky brightness (SQM radiance) and the urban emissions of light from streetlights, residential buildings and vehicles, in the city of A Coruña, estimated from the measurements carried out in the night of 27 to 28 April 2017.

The $T_k(t)$ functions corresponding to the residential buildings and vehicle lights shown in Fig. 2 are highly fluctuating in short time scales, an artifact due to the relatively small number of individual sources that can be detected within the field of view. This effect is more noticeable in the case of the vehicle headlights, since the area of the images occupied by streets and roads is relative small. The overall trend of these functions, however, is deemed representative of the actual emissions taking place at the whole city level, and can be extracted from the raw signals by low-order polynomial fits, as shown in Fig. 3.

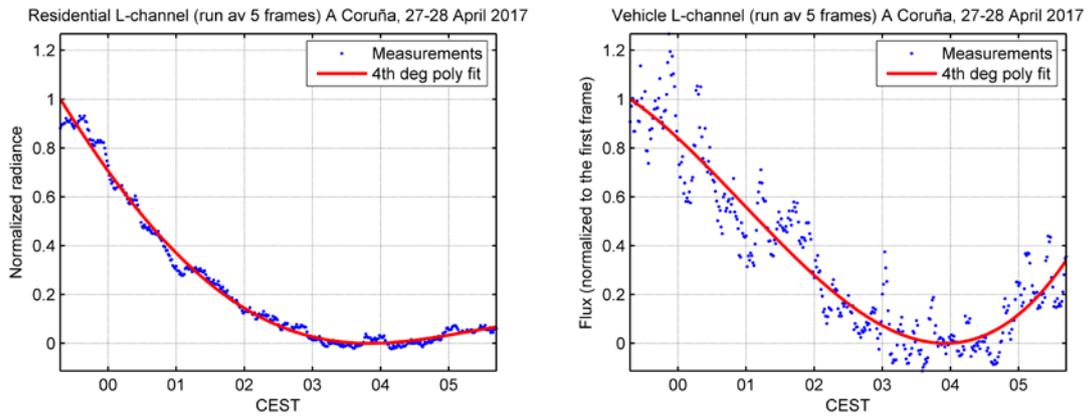

**Figure 3**: 4th degree polynomial fits of the residential and vehicle light signals shown in Fig. 3, which were used as estimates of $T_k(t)$ for these sources.

The estimated components of the modal coefficients vector are $\hat{\mathbf{c}} = (1.75, 0.49, 0.03)$ mcd/m$^2$, for streetlights, residential, and vehicle lights, respectively, being the condition number of the $\mathbf{T'T}$ matrix of order 6x10$^2$. From these components, and using Eqs. (1) and (2), the absolute and relative contribution of each type of source to the zenithal sky brightness can be estimated for every moment of the night. Figure 4 shows the results, expressed in luminance units (milicandelas per square meter, mcd/m$^2$).

The relative weight of each component of the urban emissions is shown in Figure 5. The contributions of the vehicle and residential lights diminish steadily from the end of the astronomical evening twilight until about 04:00 h (CETS), increasing continuously from that moment on until the beginning of the astronomical morning twilight, although at a slower pace. The relative weight of the constant streetlight emissions follows, as expected, the opposite trend.

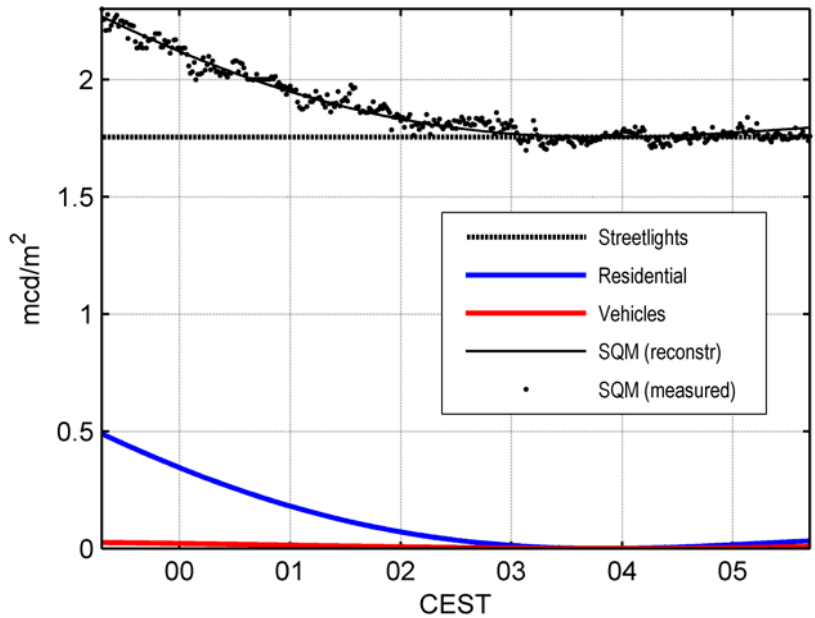

**Figure 4**: The measured SQM zenithal sky brightness of A Coruña, in absolute luminance units (mcd/m$^2$), and its streetlight, residential and vehicle components estimated by Eq. (4). The thin black line, SQM(reconstr), is the reconstructed zenithal luminance resulting from adding up these components, Eq. (1).

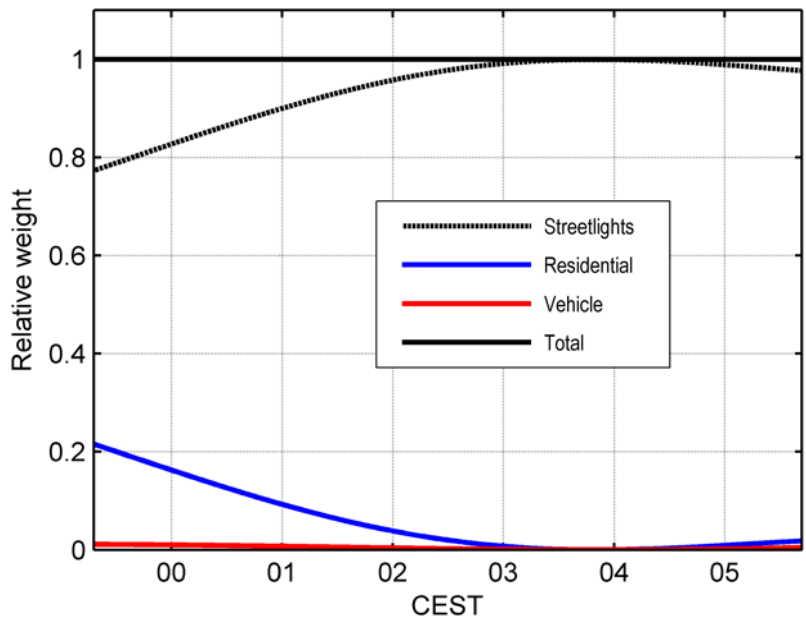

**Figure 5**: Relative weight of the components of the zenithal sky brightness throughout the night, based on the measurements carried out in A Coruña. The contributions of vehicle and residential lights diminish steadily from the end of the astronomical twilight after sunset until about 04:00 h (CEST), increasing from then on until the beginning of the astronomical twilight before sunrise. The relative weight of the constant streetlight emissions follows the opposite trend.

The maximum residential lighting contribution takes place at the beginning of the astronomical night and amounts to a 22% (0.49 mcd/m$^2$) of the total sky brightness, whilst the traffic lights seem to have a noticeably smaller contribution, on the order of 1% (0.03 mcd/m$^2$). At the same time the contribution of the constant streetlights (1.75 mcd/m$^2$) reaches its minimum relative value (77%).

*4.2. Arteixo*

The measurements in Arteixo were made in the night of 24 to 25 February, 2017, from a vantage point located at the center of the main urban nucleus, close to the rim of the large industrial park of Sabón (Fig 1, right). This site is located 10.1 km to the South West of the *Casa das Ciencias Planetarium* of A Coruña. The zenithal sky brightness measurements were made with a SQM-LU-DL detector, and the time course of the artificial light emissions was estimated from time lapse images obtained using the same camera model as in A Coruña, with an objective lens of focal length 25 mm stopped at f/4, ISO 800 and exposure time 1/25 s. One frame was taken every 2 minutes during the whole duration of the astronomical night.

As in the previous example, Figs. 6-9 show the raw $T_k(t)$ functions, the 4th-degree polynomial fits of the residential and vehicle signals, the estimated absolute contribution of each source type to the overall zenithal sky brightness (mcd/m$^2$), and the corresponding relative weights. The time course of the strong light sources present in the industrial park, clearly distinguishable from the streetlights, vehicle and residential components, was included as an independent $T_k(t)$ in the analysis. Public ornamental lights temporarily installed throughout the town, that were switched off at a precise moment of the night, were also included as an independent source term. Both the industrial and the ornamental light sources were kept essentially at constant emission levels between successive abrupt transitions to lower values, and their raw data were directly used for the analysis, without performing low-degree fits as the ones used for residential and vehicle lighting. As in the case of A Coruña, the information available *a priori* regarding the constant emission level of the public outdoor streetlights allowed to set its normalized time course function to 1, with no need of DSLR image measurement.

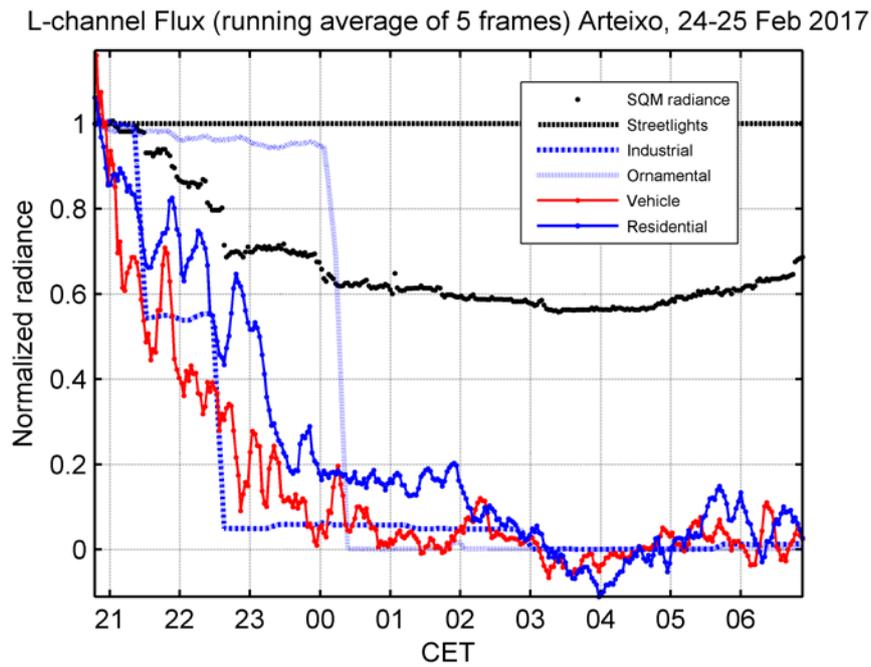

**Figure 6**: Normalized time courses, $T_k(t)$, of the zenithal night sky brightness (SQM radiance) and the urban emissions from streetlights, industrial premises, ornamental lights, vehicle, and residential buildings in the town of Arteixo, estimated from the measurements carried out in the night of 24 to 25 February 2017.

The estimated components of the modal coefficients vector are $\hat{\mathbf{c}} = $ (3.00, 0.09, 1.48, 0.82, 0.03) mcd/m$^2$, for streetlights, vehicle, residential, industrial and ornamental lights, respectively. The condition number of the $\mathbf{T'T}$ matrix is 1.5x10$^3$. These components indicate that the steady contribution of the streetlights to the zenithal sky brightness is of order 3.0 mcd/m$^2$ throughout the whole night, representing 55% of the total brightness at the beginning of the astronomical night. At the same instant the contributions of the remaining terms are at their maximum, being 2% (0.09 mcd/m$^2$) for the vehicle headlights, 27% (1.48 mcd/m$^2$) for residential lights leaked out of the windows, 15% (0.82 mcd/m$^2$) for industrial lights, and less than 1% (0.03 mcd/m$^2$) for the temporarily installed ornamental lights.

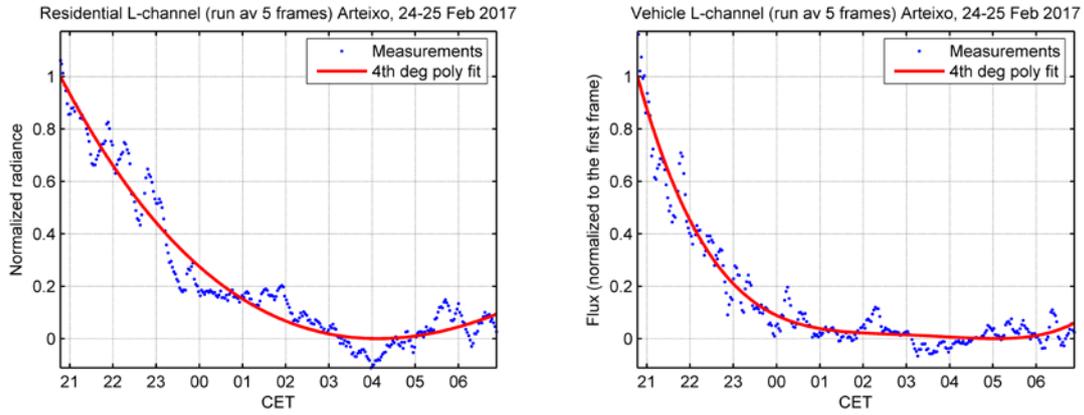

**Figure 7**: 4th degree polynomial fit of the residential and vehicle light signals shown in Fig. 6, which were used as $T_k(t)$ for these sources.

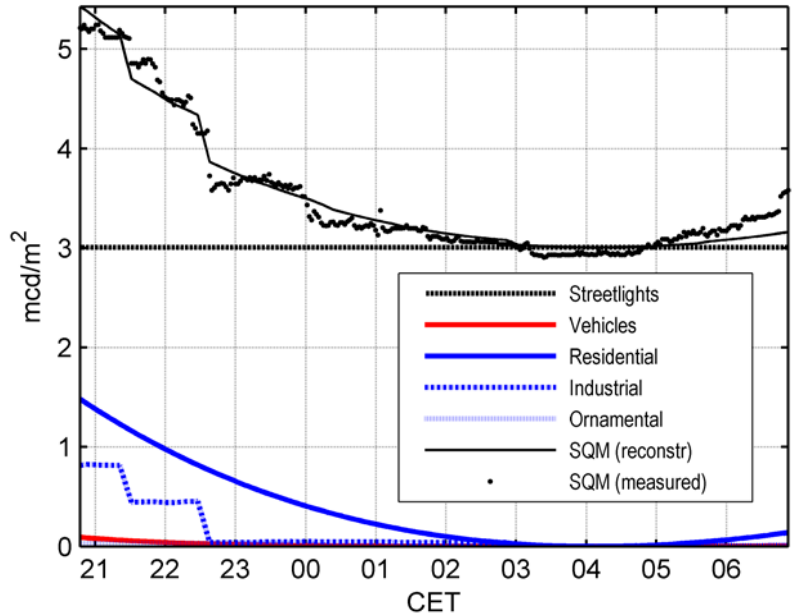

**Figure 8**: The measured SQM zenithal sky brightness of Arteixo, in absolute luminance units (mcd/m$^2$), and its streetlight, vehicles, residential, industrial and ornamental components estimated by Eq. (4). The thin black line, SQM(reconstr), is the reconstructed zenithal luminance resulting from adding up these components, Eq. (1).

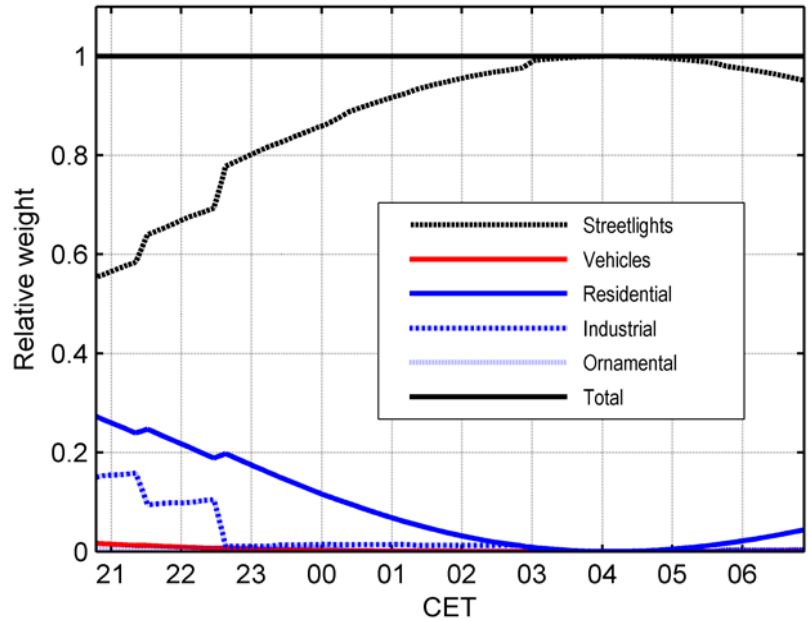

**Figure 9**: Relative weight of the components of the zenithal sky brightness troughout the night, based on the measurements carried out in Arteixo. As it was observed in A Coruña, the contributions of vehicle and residential lights diminish steadily from the end of the astronomical twilight after sunset until about 04:00 h (CEST), increasing from then on until the beginning of the astronomical morning twilight. The relative weight of the constant streetlight emissions follows the opposite trend.

## 5. Discussion

It has been previously reported that the zenithal night sky brightness in urban areas tends to decrease in the first half of the night, due to the progressive switch-off of residential, industrial and ornamental lights, and the reduction of the vehicle traffic flows. Typical darkening rates reported in worldwide studies are of order 4.5% per hour [2], that is, 0.05 mag/(arcsec$^2$·h). According to these results, the absolute difference between the brightest and the darkest point in the night may easily reach 0.3 mag/arcsec$^2$, amounting to a reduction of ~25% in the zenithal brightness of the sky. In the examples presented in Section 4, the reduction of luminance was 26% for A Coruña and 45% for the industrial site of Arteixo. Previous studies on the zenithal sky brightness of Madrid (Spain) reported a reduction of order 49%, for an estimated composition of the emissions to the upper hemisphere of 54% streetlights, 36% ornamental and commercial lights and about 9% residential lights [12]. This variability may be due to the different estimation methods, and may also reflect the specific light source composition and dynamics of different cities and towns.

All observations suggest, however, that the variable components are a relevant fraction of the overall night sky brightness. Acting on this variable part of the anthropogenic emissions may help to attain significant reductions of the light pollution levels, complementing the necessary decrease of the upward light emissions produced by streetlights. A key issue for planning and implementing any remediation measures is to identify the relative contribution of each type of source. Our preliminary results suggest that the light leaked out of residential building windows may have a relevant share in the total amount of urban emissions, especially in the first hours of the night. The outwards emission from an individual window can be quantitatively estimated using suitable multiple-reflection models of indoor lighting [37-38]. A judicious use of blinds and curtains at times when looking outdoors is not required nor desired may significantly alleviate the overall light emissions from residential buildings. In contrast, the contribution of vehicle lights to the overall sky brightness seems to be significantly smaller.

The results presented in Section 4 should be taken as an illustration of the application of the proposed method, rather than as final and sharply-cut statement on the relative weight of the different types of artificial sources in the urban sky brightness. Both the method itself and the particular examples reported here suffer from some limitations that deserve further consideration. The core of the method is the reduction of the dimensionality of the description of the zenithal sky brightness, achieved by passing from a time-based representation $B(t)$ to a modal coefficients-based representation $c_i$, in terms of the time signatures of each contributing type of source $T_i(t)$, and the subsequent estimation of the modal coefficients from a sufficiently large discrete set of measurements made at times $t_s$, $s=1,...,N$. The main problem associated with this approach, from a basic standpoint, is the lack of orthogonality of the time signature functions, in the sense that $\sum_{s=1}^{N} T_i(t_s)T_k(t_s) \neq K_i \delta_{ik}$, being $K_i$ a real positive number and $\delta_{ik}$ the Kronecker-delta function. This lack of orthogonality forces the solutions of Eq. (4) to be highly dependent on the number and type of terms included in the sky brightness description, Eq. (1). An accurate estimation of the relative brightness contributions is therefore contingent upon the correct identification of all relevant types of contributing sources. Besides, this lack of orthogonality, especially noticeable for the residential and vehicle lighting contributions due to the similarity of their time courses, puts stronger demands on the numerical precision required for an accurate solution of Eq. (4), and may help to increase the noise propagation.

From an experimental viewpoint, the main requirement associated with this approach is the need of finding a suitable vantage point to determine with sufficient accuracy and precision the time course of the relevant artificial sources. This is particularly true of the vehicle lights signal, since the overall field of view corresponding to the streets and roads in the urban nightscape imagery is generally smaller than the one corresponding to windows, and the traffic flow at the street scale (not so at the metropolitan area scale) is highly fluctuating in time. In our present study the time-lapse images were taken from a single location, trying to encompass all kinds of relevant light sources: a straightforward improvement would be to acquire simultaneous time-lapse images from a variety of urban locations, each one containing a sufficiently large sample of individual sources of a given kind. Since the zenithal sky brightness responds to the overall emissions of the urban area, the key issue is finding locations sufficiently representative of the average behaviour of the city at large.

In this work we did not address the spectral composition of the light scattered back from urban skies to the observer. The spectral composition of the light emitted by urban sources can be determined by using hyperspectral imagery, either ground-based or airborne (see [39-40] and references contained therein).

Putting aside these limitations, the results presented in this paper show the feasibility of the proposed approach, and suggest the relevance of residential lights in the build up of the overall levels of artificial skyglow in urban areas. Further observational campaigns are necessary to confirm or otherwise contradict this finding.

## 6. Conclusions

The zenithal night sky brightness in urban areas varies throughout the night following the changing time course of the anthropogenic emissions of light. Different types of artificial sources (e.g. streetlights, residential, and vehicle lights) show specific time signatures, and this feature allows for an estimation of the amount of sky brightness contributed by each of them. The method described in this paper consists on reducing drastically the dimensionality of the problem by transforming the time-based representation of the zenithal sky brightness into a modal coefficients' one, in terms of the time course signature of each source type. The modal coefficients, and hence the absolute and relative contributions of each light source, can be estimated by linear least-squares fits of the modal expansion to the measured values of the zenithal sky brightness. The zenithal sky brightness may be monitored using conventional low-cost light meters. The time course of the sources can be determined from wide-field time-lapse photometry. The preliminary results shown in this paper suggest that artificial light leaking out of the windows of residential buildings accounts for a significant share of the varying part of the zenithal sky brightness, whilst the contribution of vehicle lights is significantly smaller. Further studies are required to confirm this point.

**Acknowledgments**

This work was developed within the framework of the Spanish Network for Light Pollution Studies, REECL (AYA2015-71542-REDT). Thanks are given to *Casa das Ciencias Planetarium* of A Coruña for the SQM data of the night of 27 to 28 April, 2017, and to Mrs V. Rivas for the kind assignment of her office at the *ETS Naútica y Máquinas* as a vantage point for recording the time-lapses in A Coruña. CITEUC is funded by National Funds through FCT - Foundation for Science and Technology (project: UID/Multi/00611/2013) and FEDER - European Regional Development Fund through COMPETE 2020 Operational Programme Competitiveness and Internationalization (project: POCI-01-0145-FEDER-006922).

**APPENDIX**

Some formal details for arriving at Eq. (1) are described here. Let $L(\lambda;\boldsymbol{\alpha},t)$ be the spectral radiance of the sky at time $t$ in the direction $\boldsymbol{\alpha}$ (a two-dimensional vector whose components are e.g. the altitude and azimuth of the points of the sky measured in the observer reference frame) and wavelength $\lambda$, expressed in units $\text{Wm}^{-2}\text{sr}^{-1}\text{nm}^{-1}$. The artificial component of this spectral radiance is the sum of the sky radiances produced by all individual artificial light sources present in the region surrounding the observer. If $M$ different kinds of light sources $L_i(\lambda;\boldsymbol{\alpha},t)$, $i=1,...,M$, can be identified (e.g. streetlights, residential, vehicle lights,...) each one being characterized by a particular aggregate spectral composition and a specific time course, the overall radiance can be written as:

$$L(\lambda;\boldsymbol{\alpha},t) = \sum_{i=1}^{M} L_i(\lambda;\boldsymbol{\alpha},t). \tag{A1}$$

If the atmospheric conditions remain stable throughout the night (or, as a less stringent condition, if the changing atmospheric conditions affect in the same way to the fraction of light scattered towards the observer for all types of light sources), and if the aggregated angular emission pattern of each type of source does not vary significantly throughout the night (a reasonable assumption, due to the high number of individual sources with random orientations that build up the emission of each source class), the sky radiances $L_i(\lambda;\boldsymbol{\alpha},t)$ will be proportional to the aggregated emitted spectral density flux, $\Phi_i(\lambda;t)$ (W·nm$^{-1}$), at each instant $t$. Under these assumptions, the coefficient of proportionality, $g_i(\lambda,\boldsymbol{\alpha})$, that depends on the kind of source ($i$), the wavelength ($\lambda$) and the direction in the sky ($\boldsymbol{\alpha}$), will not depend on time. We have, then:

$$L_i(\lambda;\boldsymbol{\alpha},t) = g_i(\lambda,\boldsymbol{\alpha})\Phi_i(\lambda;t). \tag{A2}$$

On the other hand the measured brightness, $B(t)$, is the result of integrating the radiance $L(\lambda;\boldsymbol{\alpha},t)$ over the field of view of the detector, characterized by the weighting function $F(\mathbf{\alpha})$, and subsequently integrating the result over wavelengths, weighted by the spectral sensitivity of the detector, $V(\lambda)$. The (noiseless) brightness measurements can hence be described by:

$$B(t) = \int_\Omega \int_0^\infty L(\lambda;\mathbf{\alpha},t) F(\mathbf{\alpha}) V(\lambda) \mathrm{d}\lambda \, \mathrm{d}^2\mathbf{\alpha} = \sum_{i=1}^M \int_\Omega \int_0^\infty L_i(\lambda;\mathbf{\alpha},t) F(\mathbf{\alpha}) V(\lambda) \mathrm{d}\lambda \, \mathrm{d}^2\mathbf{\alpha}, \quad (A3)$$

where $\mathrm{d}^2\mathbf{\alpha} = \cos(h)\mathrm{d}h\mathrm{d}\phi$, with $h$ standing for altitude and $\phi$ for azimuth, is the solid angle element. The solid angle integral is formally carried out over the celestial hemisphere $\Omega$, spanning $2\pi$ sr. Substituting Eq. (A2) into Eq. (A3) we have:

$$B(t) = \sum_{i=1}^M \int_\Omega \int_0^\infty g_i(\lambda,\mathbf{\alpha}) \Phi_i(\lambda;t) F(\mathbf{\alpha}) V(\lambda) \mathrm{d}\lambda \, \mathrm{d}^2\mathbf{\alpha}, \quad (A4)$$

and, after integrating over the angular variables:

$$B(t) = \sum_{i=1}^M \int_0^\infty G_i(\lambda) V(\lambda) \Phi_i(\lambda;t) \mathrm{d}\lambda, \quad (A5)$$

where $G_i(\lambda) = \int_\Omega g_i(\lambda,\mathbf{\alpha}) F(\mathbf{\alpha}) \mathrm{d}^2\mathbf{\alpha}$ is the weighted integral of $g_i(\lambda,\mathbf{\alpha})$ over the $2\pi$ sr celestial hemisphere above the observer.

Finally, if the spectral composition of the radiance emitted by each type of source does not vary throughout the night (note that the different types of sources are defined, among other criteria, by this condition), the aggregated spectral radiant flux can be factored out as:

$$\Phi_i(\lambda;t) = \tilde{\Phi}_i(\lambda) T_i(t) \quad (A6)$$

where $T_i(t)$ is the function describing the time course of the emissions and $\tilde{\Phi}_i(\lambda)$ is the spectral composition of the *i*-th type of source, which, by assumption, does not depend on time. The scaling and dimensions of $T_i(t)$ and $\tilde{\Phi}_i(\lambda)$ can be assigned arbitrarily, provided that their product gives the correct value and units for $\Phi_i(\lambda;t)$. A useful choice is to identify $\tilde{\Phi}_i(\lambda)$ with the overall spectral flux (W·nm$^{-1}$) emitted at a given time $t_0$ of the night, $\tilde{\Phi}_i(\lambda) \equiv \Phi_i(\lambda;t_0)$, in which case the time course signature $T_i(t)$ turns out to be dimensionless and normalized such that $T_i(t_0) = 1$.

From Eqs. (A5) and (A6) we can write:

$$B(t) = \sum_{i=1}^{M} \left[ \int_0^\infty C_i(\lambda) V(\lambda) \tilde{\Phi}_i(\lambda) \mathrm{d}\lambda \right] T_i(t), \tag{A7}$$

and hence we get Eq. (1)

$$B(t) = \sum_{i=1}^{M} c_i T_i(t), \tag{A8}$$

where

$$c_i = \int_0^\infty C_i(\lambda) V(\lambda) \tilde{\Phi}_i(\lambda) \mathrm{d}\lambda \tag{A9}$$

are constant coefficients that do not depend on time, and whose units are the same as the units of $B(t)$.

According to the recommended SI practice, $B(t)$ is usually expressed as a weighted radiance (Wm$^{-2}$sr$^{-1}$), specifying the photometric band $V(\lambda)$ used for the measurements. Note that when this band is coincident with the photopic (or scotopic) CIE spectral efficacy function of the human visual system, the brightness can be equivalently described in the SI luminous units cd/m².